\documentclass[aps,pre,twocolumn,superscriptaddress,showpacs,preprintnumbers,floatfix]{revtex4-2}

\usepackage{hyperref}
\usepackage{natbib}
\usepackage{etoolbox}
\apptocmd{\thebibliography}{\raggedright}{}{}

\usepackage[utf8]{inputenc}
\usepackage{subfiles}
\usepackage{amsfonts}
\usepackage{graphicx}
\usepackage{physics}
\usepackage{float}
\usepackage{amsmath}
\usepackage{mathtools}
\usepackage[english]{babel}
\usepackage{csquotes}

\usepackage{tcolorbox}

\newcommand{\Ck}{C_{\mathcal{K}}} 
\newcommand{\U}{\mathcal{U}} 
\newcommand{\PF}{\mathcal{P}} 
\newcommand{\kr}{\kappa} 
\newcommand{\M}{\mathcal{M}} 

\newcommand{\hrho}{\hat{\rho}} 
\newcommand{\hkr}{\hat{\kappa}} 

\newcommand{\sbraket}[2]{\left(#1|#2\right)}

\newcommand{\clim}{\approx}

\newcommand{\bnt}[2]{\beta_{#1\,#2}}
\newcommand{\atn}[2]{\alpha_{#1\,#2}}

\newcommand{\uba}{Universidad de Buenos Aires, Facultad de Ciencias Exactas y Naturales, Departamento de Física. Buenos Aires 1428, Argentina}
\newcommand{\ifiba}{CONICET - Universidad de Buenos Aires, Instituto de Física de Buenos Aires (IFIBA). Buenos Aires 1428, Argentina}

\newcommand{\acceptednote}{\textbf{Accepted Manuscript}. This is the Accepted Manuscript version of an article accepted for publication in Journal of Physics A: Mathematical and Theoretical. The Version of Record is available at \href{https://doi.org/10.1088/1751-8121/ae8a93}{https://doi.org/10.1088/1751-8121/ae8a93}. This Accepted Manuscript is available for reuse under a CC BY-NC-ND licence after the 12 month embargo period provided that all the terms and conditions of the licence are adhered to.}

\begin{document}

\title{Quantum-to-classical correspondence in Krylov complexity}
    \thanks{\acceptednote}

    \author{Gast\'on F. Scialchi}
    \email[\textbf{Corresponding author}. E--mail address:]{gscialchi@df.uba.ar}
    \affiliation{\uba}
    \affiliation{\ifiba}

    \author{Augusto J. Roncaglia}
    \affiliation{\uba}
    \affiliation{\ifiba}

    \author{Diego A. Wisniacki}
    \affiliation{\uba}
    \affiliation{\ifiba}

    \begin{abstract}
        We study quantum-to-classical correspondence of the Krylov space for evolutions driven by unitary maps with a classical limit.
        This entails a proper definition of corresponding quantum and classical operators, inner products and initial states.
        We prove that with these definitions the purely classical Krylov space is indeed obtained as the asymptotic $\hbar\to 0$ expansion of the quantum Krylov space,
        and provide several examples of such correspondence.
        We use these examples to analyze some general aspects about the evolution of the Krylov complexity
        as they relate to the phase-space representation for the Krylov states.
        Additionally, we discuss alternative definitions to obtain the correspondence and why they fail.
        This paper constitutes a first step in understanding complexity and ergodicity of unitary evolution through the Krylov perspective as they relate to classical dynamical notions.
    \end{abstract}

\maketitle

\section{Introduction}
\label{sec:introduction}
In its inception, the field of quantum chaos was concerned with manifestations of the underlying classically chaotic dynamics in the quantum regime.
Many quantities have been studied whose behavior bring to light such manifestations.
One example is the relationship between the spectral statistics of a quantum system,
such as level and eigenstate statistics,
and the chaoticity of its underlying classical counterpart~\cite{PhysRevLett.52.1, haake1991quantum, mehta2004random, wimberger2014nonlinear, PhysRevA.42.1013, MKus_1988, izrailev1987chaotic}.
This statistical approach probes global and time-independent properties of the system.
A complementary dynamical picture has also been developed,
where the focus is on the study of time-dependent quantities
whose behavior show signatures of the related classical dynamics.
Examples of these quantities are the Loschmidt echo \cite{Wisniacki2012, PhysRevLett.86.2490},
as a probe of the sensitivity to perturbations in the time-reversed evolution of a system,
and the out-of-time-order correlator (OTOC),
whose growth is associated with the spread of quantum information \cite{shenker2015stringy}
and has also been related with the Lyapunov exponent if an underlying classical system is present \cite{swingle2018unscrambling,Garcia-Mata:2023,PhysRevLett.121.210601}.
In quantum systems where a classical limit may not exist,
its chaoticity is \textit{defined} according to how the aforementioned quantities behave.

Krylov complexity is a recent entry in the dictionary of dynamical quantum chaos.
Its original formulation arises in the context of operator growth,
where an initially ``simple" operator becomes increasingly ``complex" as it evolves in the Heisenberg picture,
in the sense that it requires an increasing amount of elements from a reference operator basis to construct it,
in this case the Krylov basis.
A ``universal operator growth hypothesis" implies that under chaotic Hamiltonian dynamics Krylov complexity grows exponentially with a rate that upper-bounds that of other operator growth measures such as the OTOC \cite{Parker2019},
which itself reflects aspects of the underlying classical dynamics (if it exists) as its growth-rate is related to the Lyapunov exponent,
like mentioned above.
Under general unitary quantum dynamics,
where there may not exist an underlying static Hamiltonian driving the evolution,
a complementary picture has been proposed:
for chaotic systems, the presence of a ``maximally ergodic regime" is a universal feature
which in turn implies maximal growth of Krylov complexity \cite{PhysRevB.111.014309}.

Although the Krylov picture for Hamiltonian dynamics has long been applied equally in both quantum and classical settings,
it was mostly used as a tool to probe close-to-equilibrium properties from the dynamics of many-body systems through the recursion method \cite{viswanath2008recursion}.
The method consists of an iterative \textit{orthogonal} expansion for the evolution of some observable of interest,
such that the resulting expansion coefficients encode specific information about the dynamics
(for example bandwidths, gap sizes, decay rates, etc.).
This requires a few ingredients: an observable, an inner product and a generator for the time evolution,
which may be the Hamiltonian or the Liouvillian in its original context~\cite{Balasubramanian_2022, Rabinovici2021, barbon2019evolution, PhysRevE.107.024217, Rabinovici_2022_integrability},
or a unitary evolution (super)operator in more general unitary evolutions~\cite{PhysRevB.110.155109, jp5f-wpkq, PhysRevE.108.054222, PhysRevE.111.L052202}
(although general non-unitary evolutions may be considered as well~\cite{bhattacharya2022operator}).
In this context, both the universal operator growth hypothesis and the maximally ergodic regime may be interpreted as statements about universal properties the expansion coefficients possess under chaotic dynamics.
It is also worth noting that the universal operator growth hypothesis was proposed \textit{including} purely classical dynamics.
However, most of the effort since has been focused on the study of systems in the quantum or semiclassical regime.

The aim of this line of work is to establish an understanding on the notions of integrability and chaos as viewed from the Krylov perspective
in terms of classical dynamical concepts.
As a first step, it is necessary to understand how to correctly define the construction
(i.e., the \textit{ingredients} mentioned earlier)
of both a quantum and a classical Krylov space such that they obey the correspondence principle.
This is the particular goal of this paper,
with a focus in evolutions described by general unitary (super)operators.
More precisely, we proved that
under suitable choices of inner products and initial states, the resulting Krylov spaces for
the quantum evolution of a density matrix
and the classical evolution of a probability distribution in phase space, respectively,
become identical in the classical limit.
We do this by considering a phase space representation of the quantum Krylov states through quasiprobability distributions,
and showing that in the limit $\hbar\to 0$ these distributions match the classical ones exactly.
We show examples of the correspondence in action in integrable systems with a weak non-linearity,
where the Ehrenfest time is sufficiently long to make a clear comparison between the quantum and classical.
Although we do not delve into chaotic dynamics here,
the conceptual framework established is nonetheless general.
Our particular focus on density matrices may also be relaxed,
as the results are readily extended to general observables.
Additionally,
we found that an alternative approach to obtain the classical Krylov space in correspondence with a given quantum one
based on choosing an initial classical distribution that matches the quasiprobability distribution of the initial quantum state
does not work.

The paper is structured as follows.
In Sec.~\ref{sec:arnoldi} we review the construction of Krylov space for unitary evolution through the Arnoldi iteration,
and define both the quantum and classical Krylov spaces and the Krylov complexity.
We then show in Sec.~\ref{sec:correspondence} that these definitions obey a quantum-to-classical correspondence through a phase-space representation via a quasiprobability distribution.
In Appendix~\ref{sec:ap:lack_correspondence} we additionally study an alternative approach to correspondence:
to match the classical distribution to that of the phase-space representation for the quantum system,
and discuss why it fails.
Examples of the correspondence in action are shown in Sec.~\ref{sec:examples},
where we compare the Krylov spaces and complexity through the quantum-to-classical transition in quantum models with their purely classical counterparts,
with both linear and non-linear dynamics.
These examples also serve to interpret general aspects of Krylov space in terms of the phase-space picture.
We conclude in Sec.~\ref{sec:conclusions} with a summary and some final remarks.


\section{Quantum and classical Krylov space}
\label{sec:krylov_space}
Krylov subspace methods were originally developed as efficient iterative ways to solve high-dimensional sparse linear systems and eigenvalue problems,
with prominent examples including the Arnoldi iteration~\cite{arnoldi1951principle} and the Lanczos algorithm~\cite{parlett1998symmetric}.
In a physical setting, both construct the Krylov subspace spanned by the evolution generated through repeated applications of some generator $G$ onto an initial state or observable $\psi$,
and yield an orthonormal \textit{Krylov basis} adapted to the dynamics (which requires the definition of an inner product).
This process is rather generic and may be applied to both Schrödinger or Heisenberg evolutions,
with $G$ representing a Hamiltonian, Liouvillian, unitary evolution (super)operator, or more general evolution operators.
This framework has been applied to both closed and open quantum systems, including driven~\cite{PhysRevE.108.054222, PhysRevE.111.L052202}, dissipative~\cite{bhattacharya2022operator} and even non-hermitian dynamics~\cite{PhysRevB.111.174207}.
The relationship between the Krylov complexity of density matrices and states has also been explored~\cite{NANDY20251}.
It can also be applied in a classical setting, where $G$ may represent the classical Liouvillian and $\psi$ a phase-space function~\cite{viswanath2008recursion,Parker2019,das2026integrabilitybreakingsemiclassicalstrings}.
Various quantities derived from the Krylov construction can then be used to characterize dynamical properties such as transport, operator growth, complexity, and chaos~\cite{Rabinovici_2022_localization, Rabinovici_2022_integrability, PhysRevE.107.024217, PhysRevE.109.054209, PhysRevE.111.014220, PhysRevE.111.L052202, Erdmenger2023, NANDY20251}.

In this work, we focus our effort on the dynamics of observables driven by unitary (super)operators.
This choice is motivated by the fact that classical area-preserving maps are the simplest examples of systems exhibiting both regular and chaotic motion, and,
upon quantization, their dynamics are described by unitary Floquet operators.
On the other hand, the choice on dynamics of observables rather than wavefunctions stems from the former admitting a more direct correspondence to classical phase-space functions.
Accordingly, we construct the Krylov space through the Arnoldi iteration, which is a formulation naturally suited to dynamics generated by a unitary propagator,
whereas the Lanczos algorithm finds its utility by exploiting the hermitian structure of Hamiltonian generators.

\subsection{Arnoldi iteration construction}
\label{sec:arnoldi}
Both quantum and classical dynamics can be formulated in terms of an operator $\U$ that propagates an ``observable" $\rho_0$ forwards in time,
which in a quantum setting refers to a hermitian operator acting on a Hilbert space
and classically to a function defined in phase space.
This fact means that in both cases the Krylov space can be constructed much the same way:
by considering the repeated applications of the propagator and obtaining an orthonormal basis for the space that it spans (the Krylov basis).
This requires the definition of an inner product that determines what ``orthonormal" means,
and a procedure to construct the aforementioned basis,
the Arnoldi iteration is one such procedure that we detail below.

In both quantum and classical dynamics,
start with an initial observable $\rho_0$ and a propagator $\U$ that evolves it in one time step $\rho_t = \U \rho_{t-1}$,
where $\rho_t = \U^t \rho_0$ with $t\in \mathbb N_0$.
Define an inner product $\sbraket{\cdot}{\cdot}$ and induced norm $\norm{\cdot} = \sqrt{\sbraket{\cdot}{\cdot}}$.
Then, let $\kr_0 \equiv \rho_0/\norm{\rho_0}$ be the first element of the Krylov basis;
the rest is obtained recursively by the Arnoldi iteration
\begin{equation}
    b_n \kr_n = \U \kr_{n-1} - \sum_{l=0}^{n-1} \sbraket{\kr_l}{\U\kr_{n-1}} \kr_l
    \label{eq:arnoldi},
\end{equation}
yielding the Krylov basis $\{\kr_0, \kr_1, ...\}$ with $\sbraket{\kr_n}{\kr_m} = \delta_{nm}$.
The exact same basis can be obtained by applying a Gram-Schmidt orthonormalization directly to the sequence $\{\rho_0, \rho_1, ...\}$ as detailed in Appendix~\ref{sec:ap:gram-schmidt}.
This alternative is useful when one doesn't have explicit access to the propagator $\U$ itself,
but only to the aforementioned sequence,
as will be the case in the classical setting.
As a practical note, we carry out the orthonormalization step twice
(a modification usually referred to as ``reorthonormalization"),
be it in the Arnoldi iteration or the Gram-Schmidt procedure.
This mitigates the possible loss of orthogonality in the resulting basis as a consequence of finite-precision numerics.

In the resulting Krylov basis,
the propagator is represented by a matrix with an upper Hessenberg form,
and is completely parametrized by the Arnoldi sequences~\cite{PhysRevB.111.014309}
\begin{eqnarray}
    b_n = \sbraket{\kr_n}{\U\kr_{n-1}}, \text{ }
    a_n = \sbraket{\kr_n}{\U\kr_n}, \text{ }
    c_n = \sbraket{\kr_0}{\U\kr_n}.
    \label{eq:arnoldi_sequences}
\end{eqnarray}
Its lower diagonal is defined by $b_n$, and the rest of the non-zero elements by
\begin{equation}
    \U_{m n} = \sbraket{\kr_m}{\U\kr_n} = \frac{a_m}{c_m} c_n \quad \forall\, m\leq n
    \label{eq:propagator_sequences}.
\end{equation}

One may then expand this basis in terms of the evolution
\begin{equation}
    \kr_n = \sum_{t=0}^n \atn{t}{n}\,\rho_t,
    \label{eq:kn_rhot}
\end{equation}
or conversely
\begin{equation}
    \rho_t = \sum_{n=0}^t \bnt{n}{t}\,\kr_{n}
    \label{eq:rhot_kn}.
\end{equation}
Orthogonality of the Krylov states implies that
\begin{equation}
    \sbraket{\kr_n}{\kr_m} = \sum_{t=0}^{m} \bnt{n}{t}\atn{t}{m} = \delta_{nm}
    \label{eq:kn_ortho_bnt_atn},
\end{equation}
which means that $\beta$ and $\alpha$ are recursively inverse matrices,
i.e., let $\beta^{(l)}$ and $\alpha^{(l)}$ be the $l\times l$ matrices with elements
$(\beta^{(l)})_{nt} = \bnt{n}{t}$ and $(\alpha^{(l)})_{tm} = \atn{t}{m}$, respectively,
then $\beta^{(l)} \alpha^{(l)} = \mathbb{I}_{l\times l}$ $\forall l$.
In terms of these quantities, the matrix~\eqref{eq:propagator_sequences} can alternatively be expressed as
\begin{equation}
    \U_{m n} = \sum_{t=0}^{m} \bnt{n}{t+1}\atn{t}{m}
    \label{eq:propagator_wavefunctions}.
\end{equation}
This identity is obtained by using Eq.~\eqref{eq:kn_rhot} in the right Krylov state of $\sbraket{\kr_m}{\U\kr_n}$ and that $\U\rho_t=\rho_{t+1}$.
These last two results will be particularly useful in the classical setting.

We refer to the quantity $\bnt{n}{t}$ as the \textit{wavefunction} in Krylov space,
since the system is mapped to a quasi one-dimensional hopping problem
whose evolution is given by
\begin{equation}
    \bnt{n}{t+1} = b_n \bnt{n-1}{t} + a_n \bnt{n}{t} + \atn{0}{n}^*\sum_{\nu>n}^{t} c_\nu \bnt{\nu}{t}
    \label{eq:beta_evo},
\end{equation}
where each site on the chain represents a Krylov state.
At the first site the wavefunction obeys $\bnt{0}{t} = \sbraket{\rho_0}{\rho_t}/\norm{\rho_0}^2$,
i.e., it is the normalized autocorrelation.

As stated, the above procedure is completely general and can be applied to any initial observable $\rho_0$.
Here we are interested in the evolution itself, as such our object of interest is the system's state.
In a quantum setting this means that $\rho$ represents a density matrix and
$\U$ is the unitary evolution superoperator.
In the case studies of this work, we may define it through $\U \hrho_t = U \hrho_{t-1} U^\dagger$,
with $U$ being a ``usual" unitary evolution operator
(from now on we will explicitly denote quantum observables with a hat $\hat{A}$ where sensible).
Finally, we define an operator inner product as
\begin{equation}
    \sbraket{\hrho}{\hat{\sigma}}_Q = \frac{1}{2\pi\hbar}\Trace(\hrho^\dagger \hat{\sigma})
    \label{eq:product_q}.
\end{equation}
In a classical setting, $\rho$ represents a probability distribution over phase space
and $\U$ the Perron-Frobenius operator.
From now on we will refer to this classical propagator as $\PF$.
When associated to a classical area-preserving map $\M$
its action can formally be written as
\begin{equation}
    \rho_{t+1}(x) = \PF \rho_{t}(x) = \int dx' \, \delta(x - \M(x')) \rho_{t}(x')
    \label{eq:perron_frobenius},
\end{equation}
where $x = (q, p)$ is a point in phase space.
In this case, we use the $L^2$ inner product
\begin{equation}
    \sbraket{\rho}{\sigma}_C = \int dx \, \rho(x)^{*} \sigma(x)
    \label{eq:product_c}.
\end{equation}
In practice one may not have access to the Perron-Frobenius operator for the classical evolution.
Instead, we evolve the distribution with the associated area-preserving map $\M$ such that
$\rho_{t+1}(x) = \rho_{t}(\M^{-1}(x))$, as Eq.~\eqref{eq:perron_frobenius} suggests,
and apply a Gram-Schmidt orthonormalization procedure to the sequence $\{\rho_0, \rho_1, ...\}$
to obtain the Krylov basis $\{\kr_0, \kr_1, ...\}$,
as mentioned earlier.
A lack of direct access to $\PF$ means that neither the Arnoldi sequences~\eqref{eq:arnoldi_sequences}
nor the matrix~\eqref{eq:propagator_sequences} can be computed directly.
However, from the obtained quantities we can calculate $\bnt{n}{t} = \sbraket{\kr_n}{\rho_t}_C$ [Eq.~\eqref{eq:rhot_kn}],
which provides $\atn{t}{n}$ through Eq.~\eqref{eq:kn_ortho_bnt_atn} by inverting the matrix $\beta$.
This finally yields a local matrix representation of the Perron-Frobenius operator by way of Eq.~\eqref{eq:propagator_wavefunctions},
from which the Arnoldi sequences can also be extracted.
The representation is local because it is reduced to the subspace given by the evolution subject to the chosen initial condition.

Note that, since a density matrix $\rho$ is hermitian (and a probability distribution real),
the resulting quantities defined in Krylov space are also all real.
However, the space of density matrices (probability distributions) is not closed upon the operations carried on during the construction of Krylov space.
This is because the linear combinations that result in \eqref{eq:kn_rhot} are not necessarily convex.
Thus, the resulting Krylov states will generally not be density matrices (probability distributions)
and will not obey the Kolmogorov axioms.

The procedure above iteratively constructs the minimal Krylov subspace
spanning the system evolution up to a given point in time.
In both quantum and classical dynamics this framework provides
a notion of evolution complexity encapsulated in the Krylov complexity \cite{Parker2019, Rabinovici2021, PhysRevB.111.014309}
\begin{equation}
    \Ck(t) = \sum_{n=0}^{t} n \abs{\bnt{n}{t}}^2
    \label{eq:k_complexity}
\end{equation}
as the average dimension of the subspace needed to effectively describe the evolution
from its beginning up to a time $t$.
It is important to mention that several definitions of Krylov complexity exist,
all with essentially the same expression~\eqref{eq:k_complexity}.
The ambiguity arises from how the Krylov space is generated:
it may be constructed for the Schrödinger evolution of a state
(in which case it is also referred to as the ``spread complexity"~\cite{Balasubramanian_2022, PhysRevE.110.034201})
or the Heisenberg evolution of an operator.
The relationship between the Krylov complexity of density matrices and the spread complexity of states has been explored in~\cite{NANDY20251, Caputa:2024vrn, 10.21468/SciPostPhys.15.3.080}.
Additionally, its behavior also depends on the properties of the (super)operator with which it is expanded.
In the present scenario, we are dealing with a unitary superoperator
for which the Krylov complexity possesses a linear maximal growth $\Ck(t) = t$
that is met when a completely new Krylov state is needed
to describe the system's evolution in each next time step, i.e., $\U \hkr_n = \hkr_{n+1}$.
This also implies that the propagator takes a purely lower diagonal form
$\sbraket{\hkr_m}{\U\hkr_n} = \delta_{m\,n+1}$.
Such a regime has been referred to as ``maximally ergodic" \cite{PhysRevB.111.014309}.

\subsection{Quantum-to-classical correspondence}
\label{sec:correspondence}
In this section we show that the Arnoldi iteration applied to the evolution of a quantum system
(with the inner product \eqref{eq:product_q})
has a purely classical counterpart defined by the iteration applied to the evolution a classical distribution in phase space
(with the inner product \eqref{eq:product_c}),
and that the respective resulting Krylov spaces become equivalent in the classical limit.
This equivalence is in the sense that the quasiprobability distributions that represent the quantum Krylov states in phase-space become identical to the purely classical Krylov distributions.

Many phase-space representations through quasiprobability distributions exist and are equivalent,
containing the full information of the quantum system's state.
Here we will make use of the Glauber-Sudarshan P-representation \cite{PhysRev.131.2766, PhysRevLett.10.277},
which exploits the overcompleteness of the coherent state basis to represent any density matrix in a ``diagonal form"
\begin{equation}
    \hrho = \int dx\,P_{\rho}(x) \ketbra{\alpha(x)}
    \label{eq:p_representation},
\end{equation}
such that the resulting distribution $P_{\rho}(x)$ is non-negative if the quantum state admits a classical limit \cite{Mandel_Wolf_1995}
(any observable $\hat{A}$ can be so represented, but generally $P_A$ won't be non-negative).
The transformation to obtain this distribution is linear,
so Eq.~\eqref{eq:kn_rhot} implies that the quantum Krylov states also admit such a representation,
although their P-distributions need not be positive as a consequence of the non-convex linear combination.
We discuss the choice of quasiprobability distribution in more detail further below.

We will now establish the quantum-to-classical correspondence.
Consider the quantum Arnoldi iteration \eqref{eq:arnoldi} arising from an initial state $\hrho$ and unitary superoperator $\U$.
By representing its Krylov states in phase space,
their P-distributions follow a similar scheme
\begin{equation}
    b_n P_{\kr_n}(x) = P_{\U \kr_{n-1}}(x) - \sum_{l=0}^{n-1} \sbraket{\hkr_l}{\U\hkr_{n-1}}_Q P_{\kr_l}(x)
    \label{eq:arnoldi_glauber}.
\end{equation}
On the other hand, consider the classical Krylov space constructed by applying the classical Arnoldi iteration to the initial classical distribution $\rho_0(x) \equiv P_{\rho_0}(x)$ with the propagator $\PF$, that is:
\begin{equation}
    b_n \kr_n(x) = \PF \kr_{n-1}(x) - \sum_{l=0}^{n-1} \sbraket{\kr_l}{\PF\kr_{n-1}}_C \kr_l(x)
    \label{eq:arnoldi_cl}.
\end{equation}
This way, since by definition the quantum and classical initial states are identical in phase space in the $\hbar\to 0$ limit,
to get \eqref{eq:arnoldi_cl} from \eqref{eq:arnoldi_glauber} it suffices
for the semiclassical quantum evolution to follow the classical~\cite{asymptoticNotation}:
\begin{equation}
    P_{\U \rho}(x) \clim P_\rho (\M^{-1}(x)) \equiv \PF P_\rho (x)
    \label{eq:evo_P_semicl};
\end{equation}
and for the quantum and classical inner products to match in the semiclassical limit,
which we may express as the following two conditions:
\begin{eqnarray}
    P_{\U \kr_n}(x) &\clim& \PF \kr_n(x)  \label{eq:cond1_PUk_PFPk},\\
    \sbraket{\hkr_l}{\hkr_m}_Q &\clim& \sbraket{\kr_l}{\kr_m}_C \label{eq:cond2_prodQ_prodC},
\end{eqnarray}
where in \eqref{eq:cond1_PUk_PFPk} we are also implicitly requiring that
\begin{equation}
    P_{\kr_{n}}(x) \clim \kr_{n}(x)
    \label{eq:cond3_Pk_k},
\end{equation}
although this is true on its own provided \eqref{eq:cond1_PUk_PFPk} and \eqref{eq:cond2_prodQ_prodC} hold up to the $(n-1)$th Krylov state.
To show that these conditions indeed hold, we will first need a few related results.

The first result concerns the semiclassical evolution generated by the quantum map.
Semiclassically, quantum maps $U$ with a classical counterpart $\M$ evolve coherent states into \textit{squeezed} coherent states centred on the classical trajectory up to an $O(\hbar^{1/2})$ correction \cite{robert2021coherent},
which in turn implies that the P-distribution of the evolved state follows Eq.~\eqref{eq:evo_P_semicl}.
The semiclassical propagation of coherent states is treated rigorously in the aforementioned reference,
but for clarity we provide in Appendix~\ref{sec:ap:semiclassical_evo_hus} an explicit proof of Eq.~\eqref{eq:evo_P_semicl} by direct calculation for a frequently studied type of quantized classical map.
The validity of this result, and of its repeated application $P_{\U^t \rho}(z) \approx \PF^t P_\rho(z)$,
depends on the nature of the initial state and classical map.
Particularly, on the stretching caused by the evolution to the initial distribution,
which is determined by the stability of the classical trajectories it has support on,
and quantified by the corresponding local Lyapunov exponents.
Semiclassical correspondence is thus expected to be valid up to the Ehrenfest time $\tau_E$,
which is controlled by said stability
such that $\tau_E \sim 1/\hbar^{1/2}$ (or some other power) for a stable trajectory
and $\tau_E \sim \abs{\log(\hbar)}/\lambda$ for an unstable one~\cite{Shepelyansky:2020, PhysRevLett.89.040403},
where $\lambda$ is the Lyapunov exponent.

The second result relates the quantum and classical inner products
[Eqs.~\eqref{eq:product_q} and \eqref{eq:product_c}, respectively]
through the P-representation \eqref{eq:p_representation}.
By using such a decomposition for any two quantum observables $\hrho$, $\hat{\sigma}$
and by expressing the trace in the coherent-state basis
$\Trace(\hat{A}) = \int dx\,\ev{\hat{A}}{\alpha(x)}/2\pi\hbar$,
their quantum inner product can be written as
\begin{multline}
    \sbraket{\hrho}{\hat{\sigma}}_Q =
    \iint dy dz P_{\rho}(y) P_{\sigma}(z) \frac{1}{2\pi\hbar} e^{-\frac{1}{2\hbar} (z-y)^2} \\
    \approx
    \int dy P_{\rho}(y) P_{\sigma}(y) \equiv \sbraket{P_\rho}{P_\sigma}_C
    \label{eq:qprod_to_cprod},
\end{multline}
where we have first used the closure relation $\int dx \ketbra{\alpha(x)}/2\pi\hbar = 1$,
that the overlap between coherent states is $\abs{\braket{\beta(y)}{\gamma(z)}}^2 = e^{-\frac{1}{2\hbar} (z-y)^2}$
and that the $0$-variance limit of a Gaussian distribution is a Dirac delta.
As previously stated, any observable can be represented in the form of Eq.~\eqref{eq:p_representation},
so the first line in \eqref{eq:qprod_to_cprod} holds.
The second line is valid as long as $P_{\rho}$ and $P_{\sigma}$ have a well-defined limit,
which is true for any observable that admits a classical limit by definition.
Thus, the quantum inner product of any two \textit{classically admisible} observables converges to the \textit{classical} inner product of their P-representations.

Having established these results, we now proceed to show that conditions
\eqref{eq:cond1_PUk_PFPk}, \eqref{eq:cond2_prodQ_prodC} and \eqref{eq:cond3_Pk_k} hold.
Because the classical initial state is defined as $\rho_0(x) \equiv P_{\rho_0}(x)$,
Eq.~\eqref{eq:evo_P_semicl} means that $P_{\rho_t}(x) \approx \rho_t(x)$ for times shorter than the Ehrenfest time $t \leq \tau_E$.
Then, the inner products of the evolved quantum and classical systems satisfy
$\sbraket{\hrho_s}{\hrho_t}_Q \clim \sbraket{\rho_s}{\rho_t}_C$ for $s,t \leq \tau_E$
due to Eq.~\eqref{eq:qprod_to_cprod}.
We now proceed iteratively.
Starting the iteration~\eqref{eq:arnoldi_glauber} at $n=1$ with
$\hkr_0 = \hrho_0/\norm{\hrho_0}_Q$
we have
\begin{align*}
    b_1 P_{\kr_1}(x) &= P_{\U \kr_0}(x) - \sbraket{\hkr_0}{\U\hkr_0}_Q P_{\kr_0}(x) \\
                  &\clim \PF \kr_0(x) - \sbraket{\kr_0}{\PF \kr_0}_C \kr_0(x),
\end{align*}
with $\kr_0(x) = \rho_0(x)/\norm{\rho_0}_C$,
and where the limit is obtained because
conditions \eqref{eq:cond2_prodQ_prodC} and \eqref{eq:cond3_Pk_k} are trivially met for $n=0$.
As a result condition \eqref{eq:cond3_Pk_k} holds for $n=1$.
Because $\hkr_1$ is a linear combination of $\hrho_0$ and $\U\hrho_0$,
it also admits an expansion in the P-representation \eqref{eq:p_representation}
(although since it isn't a convex sum $P_{\kr_1}$ need not be positively defined).
Additionally, because of \eqref{eq:qprod_to_cprod},
condition \eqref{eq:cond2_prodQ_prodC} also holds for $l, m \leq 1$.
This reasoning extends inductively such that these conditions are true up to $l, m, n \leq \tau_E$.
This establishes the quantum-to-classical correspondence of the Krylov space through the phase-space representation.

Before continuing, it is worth clarifying the role played by the P-representation in the proof above.
The essential ingredient is not the P-distribution itself, but rather the operator expansion being diagonal in the basis of coherent-state projectors.
Such an expansion allows the relevant calculations
(Eq.~\eqref{eq:qprod_to_cprod} and Appendix~\ref{sec:ap:semiclassical_evo_hus})
to be carried out in terms of coherent states, which are simple and have a clear classical limit.
In this context, the P-representation appears simply as the set of expansion coefficients of the observable in the coherent-state basis.
More generally, the P-, Wigner-, and Husimi representations may all be viewed as arising from different choices of operator basis,
yielding distinct phase-space distributions that nevertheless coincide in the classical limit~\cite{quantumOptics}.
Thus, our choice of representation should be regarded as convenient rather than canonical
since other operator bases may be used, yielding different representations.

In Appendix~\ref{sec:ap:lack_correspondence} we discuss a seemingly reasonable alternative method to study quantum-to-classical correspondence.
Instead of considering a density matrix with the form \eqref{eq:p_representation},
we take either a pure coherent-state density matrix $\hrho_0 = \ketbra{\alpha(x)}$ or a ket $\ket{\psi_0} = \ket{\alpha(x)}$ in the quantum regime
and compare with the classical for a Gaussian distribution with variance $\sigma^2 = \hbar$.
With this choice, the phase-space-represented quantum distribution and classical distribution of the initial state are identical
(however, note that this approach effectively fixes the choice of classical distribution to a Gaussian,
while in principle the correspondence argued above holds for any classical distribution).
Such an approach,
where the quantum evolution of some quasiprobability distribution over phase space is compared directly to that of a corresponding classical evolution,
has its precedents (for example, Refs.~\cite{PhysRevA.71.010101, PhysRevE.79.025203, PhysRevA.65.042113} to cite a few).
Yet, we find that correspondence of the Krylov space is not attained with such method:
in the examples shown
the Krylov complexity of the pure state is consistently larger than the classical
and that of the ket saturates early in the evolution.
In both cases there is no trend towards convergence.
In these examples such behavior can be understood from the structural differences between the classical Krylov states and the quantum ones:
in the first case, they are heavily weighted towards higher $t$ in Eq.\eqref{eq:kn_rhot}
and in the second, the quantum Krylov states escape the area covered by the evolution itself.



\section{Examples}
\label{sec:examples}
In this section we show the quantum-to-classical transition in Krylov space argued above
in a few simple examples
and discuss its structure in relation to the representation of the Krylov states in phase space
and their dynamics.
Although our work is motivated by questions of chaos, ergodicity and complexity,
the focus of this paper is the argued correspondence.
For this reason, the following examples are restricted to systems in a regular or nearly-regular regime,
where the Ehrenfest time is sufficiently long to make a clear comparison between the quantum and classical.
We leave the study of mixed or chaotic dynamics for later work.

It should also be noted that study of fully chaotic dynamics in the classical setting is numerically challenging,
since the phase-space distributions attain a fractal structure that makes the evaluation of the inner products~\eqref{eq:product_c} a difficult task.

While the P-representation is a useful tool for the proof of the previous section,
the resulting quasiprobability distribution is not well behaved
as the quantum nature of the state is reflected not only in negativities, as with the Wigner distribution,
but also with a high degree of singularity.
As mentioned, many such representations exist and are equivalent.
In particular, throughout the manuscript we will make use of the Husimi distribution~\cite{husimi, SARACENO199037, P_Leboeuf_1990}
\begin{equation}
    H_\rho(x) = \frac{1}{2\pi\hbar}\ev{\hrho}{\alpha(x)}
    \label{eq:husimi},
\end{equation}
which amounts to an $\hbar$-Gaussian smoothing of the P-distribution,
making it non-singular.
\begin{figure}[hb]
    \centering
    \includegraphics[width=\linewidth]{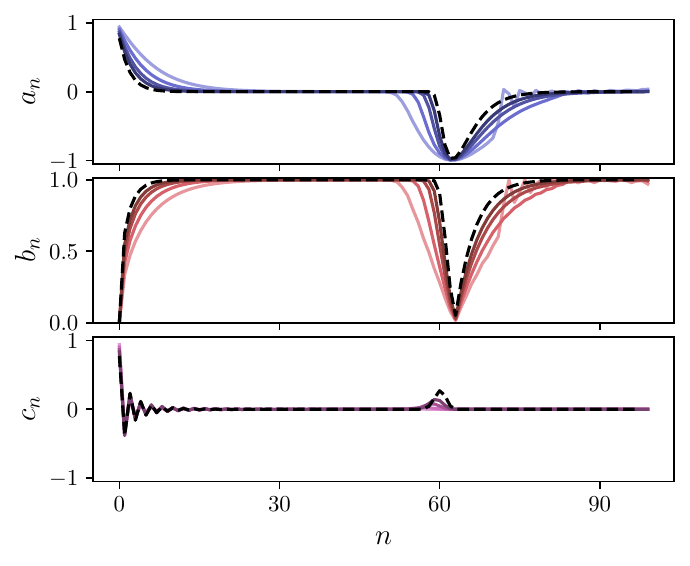}
    \caption{Quantum-to-classical correspondence in the Arnoldi sequences of the harmonic oscillator.
    The black dashed lines are the classical sequences, while
    the solid curves correspond to the quantum case with values of $\hbar \in \{2^{-4}, 2^{-5}, 2^{-6}, 2^{-7}\}$ (light to dark).}
    \label{fig:HO_sequences_correspondence}
\end{figure}

\begin{figure}[ht]
    \centering
    \includegraphics[width=\linewidth]{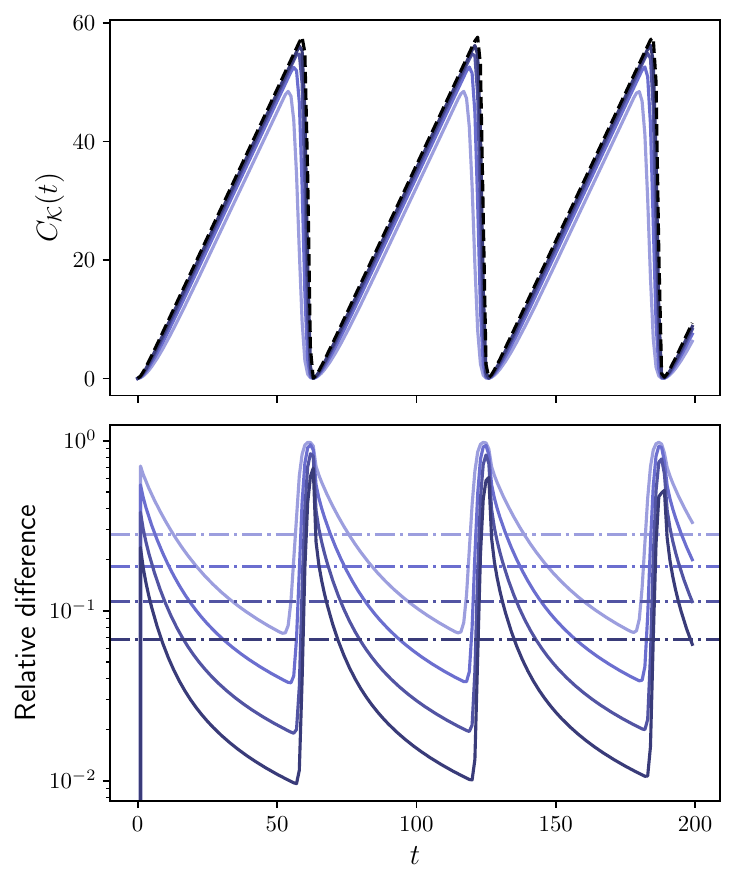}
    \caption{Quantum-to-classical correspondence in the Krylov complexity of the harmonic oscillator.
        Top panel: Krylov complexity as a function of time in the classical (black dashed line)
        and quantum (solid lines).
        Bottom panel: relative difference (to the classical) between the classical and quantum complexities (solid lines)
        and their average values (dash-dotted lines).
        The quantum curves correspond to values of $\hbar \in \{2^{-4}, 2^{-5}, 2^{-6}, 2^{-7}\}$ (light to dark).
    }
    \label{fig:HO_complexity_correspondence}
\end{figure}
\subsection{Harmonic oscillator}
The simplest case study is the Harmonic oscillator.
In this case the quantum-to-classical correspondence argued above holds trivially
since the quantum evolution itself maps coherent states onto other coherent states.

\begin{figure*}[ht]
    \centering
    \includegraphics[width=\linewidth]{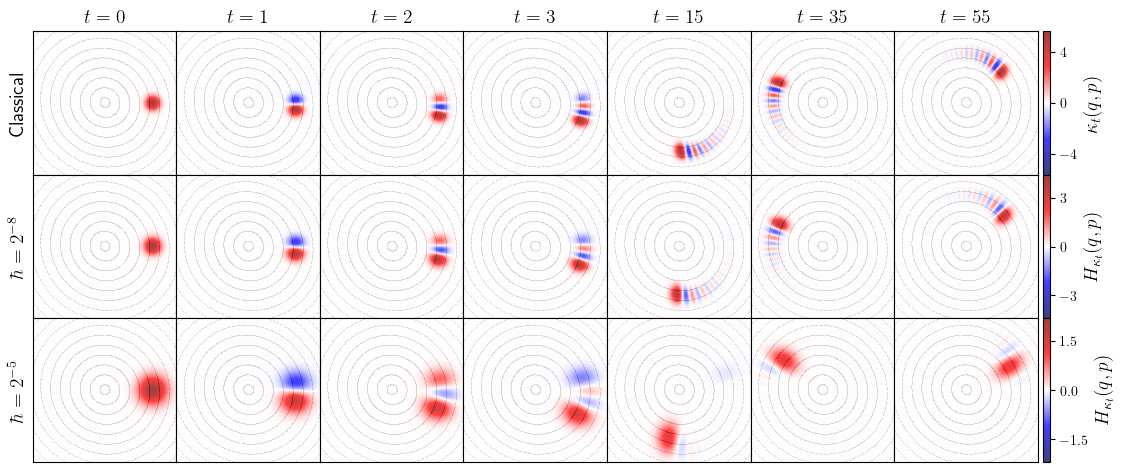}
    \caption{Classical and quantum phase space portraits of the Krylov states
        generated at various times $t$ and values of $\hbar$
        for the evolution of the harmonic oscillator.
        The quantum Krylov states are represented via the Husimi distribution [Eq.~\eqref{eq:husimi}].
    }
    \label{fig:HO_states_correspondence}
\end{figure*}
For simplicity,
we consider the Harmonic oscillator as a map through
\begin{equation}
    \begin{split}
        q' &= q + \tau p \\
        p' &= p - \tau q'
    \label{eq:ho_map},
    \end{split}
\end{equation}
having set both its frequency and mass to unity.
This is a good approximation for small values of $\tau$ and sufficient for our purposes.
Quantum mechanically we treat it with the usual unitary operator $U = e^{-i\tau(\hat{N} + \frac{1}{2})}$,
where $\hat{N}$ is the excitation number operator.
Note that having set frequency and mass to unity implies that $\hbar$ is a dimensionless quantity.
To construct the initial state we do not calculate the P-representation of a density matrix, but we do the inverse operation:
we define the classical distribution $\rho_0(q, p)$ beforehand and utilize a discrete-basis representation of the coherent states to compute $\hrho$ using Eq.~\eqref{eq:p_representation} explicitly.
In this case, the initial state chosen is defined by a Gaussian distribution
centered at a point $(q_0, p_0)$ with variance $\sigma^2$ as
\begin{equation}
    \rho_0(q, p) = \frac{1}{2\pi\sigma^2} e^{-\frac{1}{2\sigma^2} \left[(q-q_0)^2 +(p-p_0)^2 \right]}
    \label{eq:rho0_plane}.
\end{equation}
For its quantum counterpart of the form~\eqref{eq:p_representation},
the Husimi distribution~\eqref{eq:husimi} can be easily computed
\begin{equation}
    H_{\rho_0}(q, p) = \frac{1}{2\pi(\hbar + \sigma^2)}
    e^{-\frac{1}{2(\hbar + \sigma^2)}\left[(q-q_0)^2 +(p-p_0)^2 \right]}
    \label{eq:ho_hus0},
\end{equation}
which is simply the distribution arising from the direct composition of the classical ($\sigma$) and quantum ($\sqrt{\hbar}$) uncertainties,
and explicitly meets $\lim_{\hbar\to 0} H_{\rho_0}(q, p) = \rho_0(q, p)$.
For the calculations that follow we have set $\tau = 0.1$,
which sets the periodicity of the oscillator at $T\approx63$ time steps,
and parameters for the initial condition $(q_0, p_0)=(1, 0)$ and $\sigma=0.1$.

The full picture of the Krylov space and dynamics through the quantum-to-classical transition
can be seen in
Figs.~\ref{fig:HO_sequences_correspondence}, \ref{fig:HO_complexity_correspondence} and \ref{fig:HO_states_correspondence},
showing the Arnoldi sequences, the Krylov complexity and the Krylov states, respectively,
for different values of $\hbar$.
The three figures clearly show that these quantities obey the correspondence principle:
the difference between the quantum and classical Krylov complexities
and Arnoldi sequences decrease in proportion with $\hbar$
and the Krylov states become nearly identical.
It is noteworthy that in each period the complexity [Fig.~\ref{fig:HO_complexity_correspondence}]
sets into a linear growth,
indicating a ballistic propagation of the wavefunction $\bnt{n}{t}$ in completely regular dynamics.
The same growth is consistent with the profile of the Arnoldi sequences [Fig.~\ref{fig:HO_sequences_correspondence}],
which in each period quickly set into $a_n\sim 0$, $b_n\sim 1$ and $c_n\sim 0$.
In this regime Eq.~\eqref{eq:beta_evo} essentially becomes a transport equation,
so the initial wave profile given by the autocorrelation $\bnt{0}{t}$
is transported with unit speed,
and thus the Krylov complexity grows as $\Ck(t) \sim t$.
Such a regime is expected of chaotic dynamics~\cite{PhysRevB.111.014309, PhysRevE.111.014220},
although in that case it should last indefinitely, yielding a net mean growth in complexity,
which highlights the need to study the long-time behavior of the Krylov complexity.

The structure of the Krylov states [Fig.~\ref{fig:HO_states_correspondence}]
consists of a leading positive \textit{head} and a decaying sign-alternating \textit{tail}.
These features are understood by the way the orthonormalization procedure works,
which is simply to introduce negativities where the distributions overlap,
and are not a particularity of this system but universal [See App.~\ref{sec:ap:gram-schmidt}].
This feature is also reflected in the initially alternating sequence $c_n$
as they are the autocorrelation of the Krylov states~\eqref{eq:arnoldi_sequences}.

The way in which the quantum Krylov states,
as represented in phase-space by their Husimi distributions in Fig.~\ref{fig:HO_states_correspondence},
seem to simply be the classical Krylov states but for a distribution with a larger variance is deceiving.
This is indeed the case for the first one as can be deduced from Eq.~\eqref{eq:ho_hus0}, but is not true in general.
In App.~\ref{sec:ap:lack_correspondence} we show that simply matching the variance of the classical distribution to that of the Husimi won't lead to the same Krylov states nor complexity.


\subsection{Harper map}
We now turn to an example of a non-linear system
exhibiting both integrable and chaotic behavior,
although we will restrict the scope of this analysis to its regular regime
with a weak non-linearity.
The classical Harper map is defined on the unit square as
\begin{equation}
    \begin{split}
        q' &= q - k \sin{(2\pi p)} \mod 1\\
        p' &= p + k \sin{(2\pi q')} \mod 1
    \label{eq:harper_map}.
    \end{split}
\end{equation}

This system has a periodic phase space in both $q$ and $p$ directions,
implying it is contained in the unit torus.
Quantizing it leads to a Hilbert space of finite dimension $N$~\cite{P_Leboeuf_1990}
requiring the consistency relation $2\pi\hbar N = A$,
where $A$ is the area of the torus,
such that the classical limit is obtained by taking $N\to\infty$
(here we have set $A\equiv 1$, which makes $\hbar$ a dimensionless quantity).
This effectively discretizes the phase space
in the sense that the $\hat{q}$ and $\hat{p}$ operators
have a finite and discrete number of eigenstates $\ket{q_n}$ and $\ket{p_n}$, respectively.
The periodicity is reflected in the fact that a unit translation of these states must contribute with, at most, a complex phase:
\begin{eqnarray}
    \ket{q_n + 1} &= e^{-i 2\pi \bar{p}}\ket{q_n}, \, \quad \ket{p_n + 1} = e^{i 2\pi \bar{q}}\ket{p_n}
    \label{eq:floquet_phase}.
\end{eqnarray}
\begin{figure}[b]
    \centering
    \includegraphics[width=\linewidth]{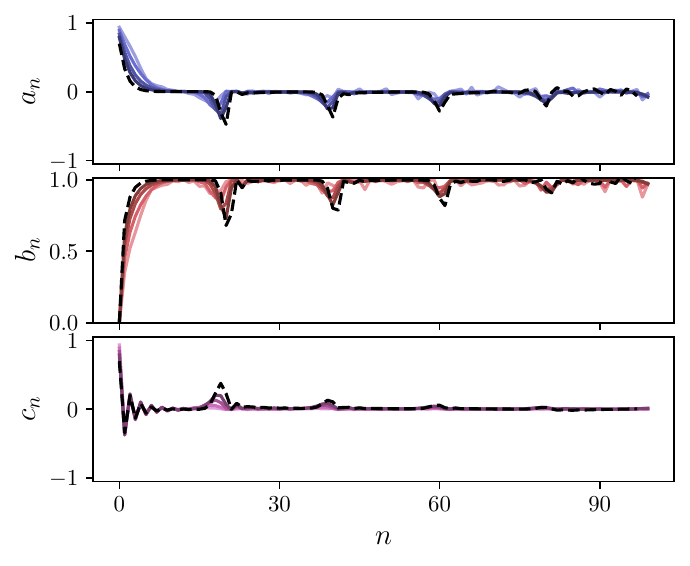}
    \caption{Quantum-to-classical correspondence in the Arnoldi sequences of the Harper map.
    The black dashed lines are the classical sequences, while
    the solid curves correspond to the quantum case with Hilbert space dimensions $N \in \{2^5, 2^6, 2^7, 2^8\}$ (light to dark).}
    \label{fig:Harper_sequences_correspondence}
\end{figure}
One then obtains the eigenvalues of the position and momentum operators as
$q_n = \frac{n+\bar{q}}{N}$ and $p_n = \frac{n+\bar{p}}{N}$, respectively, for $n=0,\ldots,N-1$.
The values $\bar{q}$ and $\bar{p}$ can take arbitrary real values between $0$ and $1$,
and are fixed: each pair $(\bar{q}, \bar{p})$ specifies a different Hilbert space~\cite{PhysRevE.65.046226, robert2021coherent}
($-2\pi \bar{p}$ and $-2\pi \bar{q}$ are called Floquet angles).
They cannot be chosen trivially as they have an effect on the dynamics:
an arbitrary choice can break symmetries present in the underlying classical map.
See Ref.~\cite{SARACENO199037} for an example in another map defined on the unit torus,
where some choices of Floquet angles would break its R-symmetry (reflection symmetry).
The Harper map possesses both an R-symmetry and T-symmetry (time reversal symmetry),
we have thus set $\bar{q} = 0.5$ and $\bar{p} = 0$
since we have observed numerically that such a choice preserves them.
Its resulting quantum evolution is described by the Floquet unitary
\begin{equation}
    U = e^{-i N k \cos(2\pi \hat{q})} e^{-i N k \cos(2\pi \hat{p})}
    \label{eq:q_harper_map}.
\end{equation}

\begin{figure}[b]
    \centering
    \includegraphics[width=\linewidth]{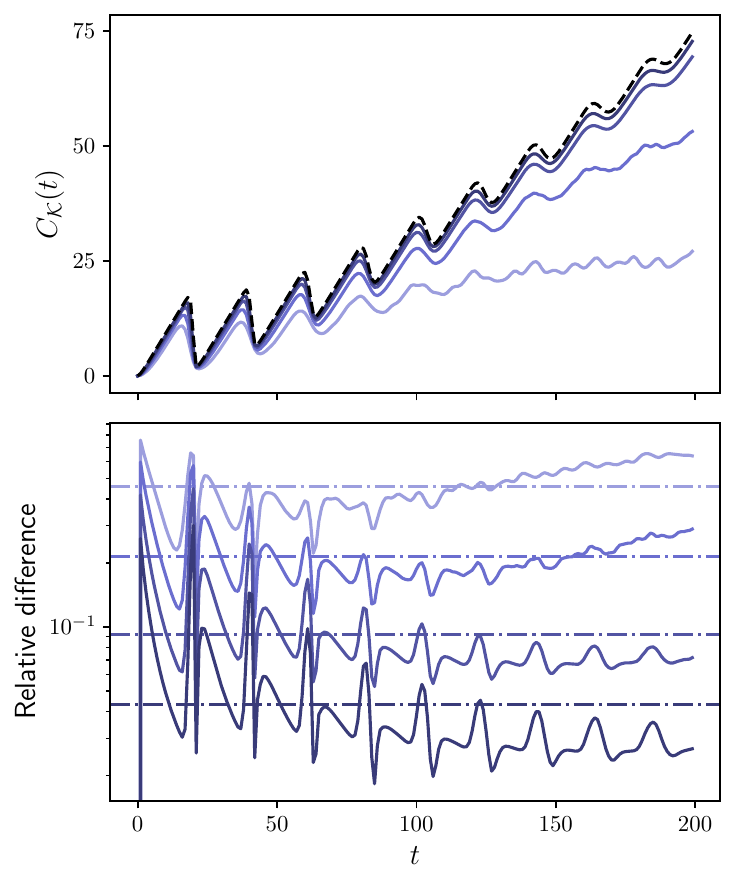}
    \caption{Quantum-to-classical correspondence in the Krylov complexity of the Harper map.
        Top panel: Krylov complexity as a function of time in the classical (black dashed line)
        and quantum (solid lines).
        Bottom panel: relative difference (to the classical) between the classical and quantum complexities (solid lines)
        and their average values (dash-dotted lines).
        The quantum curves correspond to Hilbert space dimensions $N \in \{2^5, 2^6, 2^7, 2^8\}$ (light to dark).
    }
    \label{fig:Harper_complexity_correspondence}
\end{figure}
\begin{figure*}[ht]
    \centering
    \includegraphics[width=\linewidth]{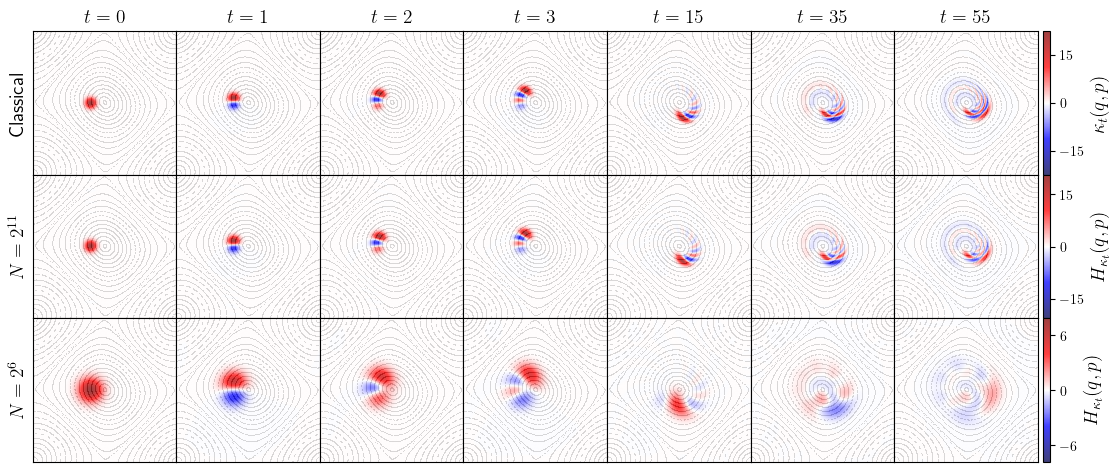}
    \caption{Classical and quantum phase space portraits of the Krylov states
        generated at various times $t$ and values of the Hilbert space dimension $N$ (where $\hbar = 1/2\pi N$)
        for the evolution of the Harper map.
        The quantum Krylov states are represented via the Husimi distribution [Eq.~\eqref{eq:husimi}].
    }
    \label{fig:Harper_states_correspondence}
\end{figure*}
The initial state here chosen is a classical Gaussian distribution as in Eq.~$\eqref{eq:rho0_plane}$,
but with periodic boundary conditions to adapt it to the unit torus
\begin{equation}
    \rho_0(q, p) =
    \frac{1}{2\pi\sigma^2} \sum_{n,m\in\mathbb{Z}}
    e^{-\frac{1}{2\sigma^2}\left[(q-q_0-n)^2 + (p-p_0-m)^2\right]}
    \label{eq:rho0_torus}.
\end{equation}
Analogously, its quantum counterpart is of the form~\eqref{eq:p_representation}.
We study this model in its regular regime $k=0.05$
with initial conditions $(q_0, p_0) = (0.4, 0.5)$ and $\sigma = 0.025$.

Figures \ref{fig:Harper_sequences_correspondence}, \ref{fig:Harper_complexity_correspondence} and \ref{fig:Harper_states_correspondence}
show the quantum-to-classical correspondence of the Krylov complexity, the Arnoldi sequences and the Krylov states, respectively,
as the dimension of the Hilbert space $N$ increases.
The observations made previously for the harmonic oscillator still generally hold,
and at early times (or values of $n$) the quantities mentioned above behave quite similarly
(as expected for the chosen initial state).
In particular,
a linear $\Ck(t) \sim t$ growth in the complexity within each cycle is also observed here.
The differences lie in the lack of periodicity given by the non-linearity.
As the distribution is pulled and stretched
new Krylov states are always being generated to encode its changing geometry,
which results in an additional net mean-growth of the complexity,
and a decaying \textit{reversion} of the Arnoldi sequences.
These two observations are understood in view of Eq.~\eqref{eq:beta_evo}:
each revival of the autocorrelation emits a new wave packet from the origin
that then propagates ballistically with unit speed.
The oscillations in complexity eventually die out
as the increasing amount of packets means the addition of another one is less impactful
(due to normalization of the wavefunction)
and the growth finally settles on a (purely) linear growth regime.

Another difference with respect to the oscillator
is seen in that the quantum complexity is ultimately bounded,
while the classical complexity seems to grow indefinitely.
The quantum complexity is trivially bounded by the dimension of the operator Hilbert space.
However, its dimension $\sim N^2$ is much larger than the saturation we observe in Fig.~\ref{fig:Harper_complexity_correspondence}.
For example, the curve with $N = 2^5$ is close to saturating at $t\sim200$, while in that case the dimension of the operator Hilbert space is $D = 1024$.



\section{Final remarks and Conclusions}
\label{sec:conclusions}
In this work we introduced a Krylov construction for classical and quantum unitary dynamics acting on observables and established an explicit quantum-to-classical correspondence between the resulting Krylov spaces through a phase-space representation.
By suitable choices of initial state and inner product, we showed that the classical Krylov space emerges as the classical limit of its quantum counterpart.
Consequently, the associated Krylov states and complexities obey the correspondence principle as well.

We presented clear examples of such correspondence in action in regular and nearly-regular systems.
In these examples, the Krylov states show a universal structure
composed of a positive leading term and a sign-alternating tail.
The time evolution of the Krylov complexity is understood in terms of the phase-space representation of the Krylov states.
The complexity grows linearly as long as new ones are being generated,
with oscillations that naturally arise from recursions, affecting its mean growth.
While the classical complexity can in principle grow indefinitely even in a bounded evolution,
we see that the quantum complexity may be ultimately capped at a value much lower than the trivial bound given by the dimension of the operator Hilbert space.
Because of the restriction to regular and near-regular dynamics,
these examples serve their purpose mostly as examples of the correspondence in action.
However,
the phase-space picture does provide an interpretation in more general regimes.
For example, in mixing systems phase-space distributions develop increasingly fine structures and a decreasing temporal overlap,
which should continuously generate new Krylov states and promote sustained complexity growth.
A detailed investigation of this scenario does however lie beyond the scope of the present work.
Facilitated by the phase-space picture,
the examples show that mere transport of the distribution can induce linear growth in complexity without the necessity of ergodicity or mixing,
as seen in Figs.~\ref{fig:HO_complexity_correspondence} and ~\ref{fig:Harper_complexity_correspondence}.
This highlights the necessity to observe the long-time behavior of the Krylov complexity.

Remarkably, in the examples presented the classical complexity seems to upper-bound the quantum.
We do not know as of yet if this is a general phenomenon, and much less so outside of the near-regular regimes here studied.
However, a heuristic argument can be made semiclassically by relying on the phase-space picture.
The quantum quasiprobability distributions typically have a larger overlap than the classical ones,
which can be seen in Eq.~\eqref{eq:qprod_to_cprod} as the Gaussian factor bridges the gap between the two $P$-distributions,
or in the Husimi distributions
(see Eq.~\eqref{eq:ho_hus0} and Figs.~\ref{fig:HO_complexity_correspondence} and ~\ref{fig:Harper_complexity_correspondence}).
This suggests a larger autocorrelation in the quantum evolution,
and consequently, a tendency towards a lower Krylov complexity.

We have also explored alternative methods in which one could propose a construction for the classical Krylov space in correspondence with a given quantum Krylov space.
Particularly by matching the classical distribution to the phase-space representation of the quantum state,
be it described by a pure density matrix or a ket.
In the examples shown,
these constructions fail to respect correspondence,
which is understood in terms of the structural differences in the resulting Krylov states.

The focus of this work has been the quantum-to-classical correspondence of Krylov constructions for unitary dynamics,
together with a phase-space interpretation of Krylov states and complexity.
For simplicity, our examples were restricted to linear and weakly nonlinear systems in the vicinity of a fixed point.
Extending these ideas to higher-dimensional, strongly nonlinear, and chaotic systems constitutes a natural next step.
While such regimes pose a numerical challenge,
particularly in the evaluation of the classical inner product for increasingly complex phase-space distributions,
the conceptual framework developed here remains applicable.

Our work grounds the Krylov construction in phase-space evolution and establishes its classical limit.
We view this as a first step toward a bridge between Krylov complexity and familiar classical dynamical notions.


\section{Acknowledgments}
This work has been partially supported by CONICET (Grant No.~PIP 11220200100568CO), UBACyT (Grant No.~20020220300049BA).

\section{Data availability}
The code from which the data that support the findings of this study are obtained is openly available at the following URL: \href{https://github.com/gscialchi/QtoC-Krylov}{https://github.com/gscialchi/QtoC-Krylov}.

\appendix

\section{Krylov space from Gram-Schmidt procedure}
\label{sec:ap:gram-schmidt}
Krylov space can equally be constructed from a direct Gram-Schmidt procedure
as well as the Arnoldi iteration \eqref{eq:arnoldi} mentioned in the main text.
In this appendix we cast the results of the construction in terms of the former.

Once defined an inner product $\sbraket{\cdot}{\cdot}$ and induced norm $\norm{\cdot} = \sqrt{\sbraket{\cdot}{\cdot}}$,
we start with the normalized temporal series $\{\rho_0, \rho_1, ...\}$.
Then, in terms of the Gram-Schmidt procedure, the iteration reads
\begin{equation}
    B_n \kr_n = \rho_n - \sum_{l=0}^{n-1} \sbraket{\kr_l}{\rho_n} \kr_l
    \label{eq:ap:gram_schmidt}.
\end{equation}
Let $G_n$ be the $n$-th Gram matrix defined as $(G_n)_{ij} = \sbraket{\rho_{i}}{\rho_{j}}$ with
$i,j=0,\ldots,n$ ($G_n$ is a $(n+1)\times(n+1)$ matrix).
In this context, and
for the inner products \eqref{eq:product_q} and \eqref{eq:product_c} considered,
it is a symmetric correlation matrix $(G_n)_{ij} = c_{\abs{j-i}} = \sbraket{\rho_i}{\rho_j}$.
The result of the iteration can then be expressed in terms of its determinants $D_n = \det(G_n)$
as the Laplace expansion
\begin{equation}
    \kr_n =
    \sum_{t=0}^n (-1)^{t+n} \frac{D_{n-1}^{(t)}}{\sqrt{D_n D_{n-1}}} \rho_t
    \label{eq:ap:kn_gs},
\end{equation}
where $D_{n-1}^{(t)}$ is the determinant of $G_n$ with its last row and its $t$-th column removed.
Since the Gram matrix is a correlation matrix, it is positive-definite and all of its determinants are positive.
Then Eq.~\eqref{eq:ap:kn_gs} implies the oscillatory nature of the Krylov states seen in the main text [Figs.~\ref{fig:HO_states_correspondence} and \ref{fig:Harper_states_correspondence}].
By comparison with Eq.~\eqref{eq:kn_rhot} it directly follows that
\begin{equation}
    \atn{t}{n} = (-1)^{t+n} \frac{D_{n-1}^{(t)}}{\sqrt{D_n D_{n-1}}}
    \label{eq:ap:alpha_gs}
\end{equation}
and then
\begin{equation}
    \bnt{n}{t} = \sum_{s=0}^n (-1)^{n+s} \frac{D_{n-1}^{(s)}}{\sqrt{D_n D_{n-1}}} c_{t-s}
    \label{eq:ap:beta_gs}.
\end{equation}
Further, from \eqref{eq:ap:gram_schmidt} and \eqref{eq:ap:beta_gs} it is easy to see that
\begin{equation}
    B_n = \bnt{n}{n} = \frac{1}{\atn{n}{n}} = \sqrt{\frac{D_n}{D_{n-1}}} > 0
    \label{eq:ap:B_n,beta,alpha,D_n}.
\end{equation}
On the other hand, one can deduce from the Arnoldi iteration that
$\bnt{n}{n} = \prod_{i=1}^n b_i$,
which relates the sequences $B_n$ and $b_n$.

\section{Semiclassical evolution of the P-distribution}
\label{sec:ap:semiclassical_evo_hus}
The quantum-to-classical correspondence argued relies on the fact that the P-distribution of the evolved quantum state asymptotically amounts to the classical evolution of the distribution itself in the semiclassical limit, as Eq.~\eqref{eq:evo_P_semicl} expresses \cite{asymptoticNotation}:
\begin{equation}
    P_{\U \rho}(x) \clim P_\rho (\M^{-1}(x)) \equiv \PF P_\rho (x)
    \label{eq:ap:evo_P_semicl}.
\end{equation}
Here we provide proof of this fact for a general class of quantized classical maps.

Instead of calculating $P_{\U\rho}(x)$ directly,
we will make use of the Husimi distribution
\begin{equation}
    H_\rho(z) = \frac{1}{2\pi\hbar}\ev{\hrho}{\alpha(z)}
    \label{eq:ap:husimi}
\end{equation}
and utilize how these two representations relate in the semiclassical limit, as follows.
Recall that the P-representation allows one to expand the density matrix in a ``diagonal" form as
\begin{equation}
    \hrho = \int dx\,P_{\rho}(x) \ketbra{\alpha(x)}
    \label{eq:ap:p_representation},
\end{equation}
such that by inserting this expression into Eq.~\eqref{eq:ap:husimi}
and taking the semiclassical limit (provided it exists) one gets
\begin{multline}
    H_\rho(z) = \frac{1}{2\pi\hbar} \int dx\,P_{\rho}(x)\abs*{\braket{\beta(x)}{\alpha(z)}}^2
    \approx
    P_{\rho}(z)
    \label{eq:ap:hus_P},
\end{multline}
where we have used
that the overlap between coherent states is $\abs{\braket{\beta(x)}{\alpha(z)}}^2 = e^{-\frac{1}{2\hbar} (z-x)^2}$
and that the $0$-variance limit of a Gaussian distribution is a Dirac delta.
This allows us to compute $H_{\U\rho}(x)$ instead of $P_{\U\rho}(x)$, which is a more straight-forward calculation.

Now the problem is shifted to that of calculating the semiclassical limit of
\begin{multline}
    H_{\U \rho}(z)
    = \frac{1}{2\pi\hbar}\ev{U\hrho U^\dagger}{\alpha(z)} \\
    = \frac{1}{2\pi\hbar} \int dw\,P_{\rho}(w) \abs*{\mel{\beta(w)}{U^\dagger}{\alpha(z)}}^2
    \label{eq:ap:husimi_Urho},
\end{multline}
where the last equality is obtained again using the P-representation for $\hrho$.
The utility of using the Husimi and P-distributions in such a way is that
it reduces the problem to that of evaluating $\mel{\beta(w)}{U^\dagger}{\alpha(z)}$ semiclasically.
The semiclassical propagation of coherent states is a topic treated rigorously and quite generally in Ref.~\cite{robert2021coherent}.
In this appendix we provide an alternative and clear result by calculating how coherent states are propagated in the semiclassical limit directly.
The essential result is that a coherent state is evolved semiclassically into a \textit{squeezed} coherent state centred on the trajectory of the classical evolution up to an $O(\sqrt{\hbar})$ correction.
As we will see, this is so because for sufficiently small values of $\hbar$ the Gaussian packet is localized enough that, while it is transported by the \textit{full} classical map,
only the \textit{linearized} classical map is involved in its deformation.
This means that the Gaussian packet evolves into another Gaussian packet, although squeezed,
with a variance proportional to $\hbar$.
As a result, the overlap in Eq.~\eqref{eq:ap:husimi_Urho} will localize the integral at the next point in the classical trajectory in the semiclassical limit.

Consider a time-dependent delta-kicked system with the Hamiltonian
\begin{equation}
    H(q, p, t) = f(q) + g(p)\sum_{n\in\mathbb{Z}} \delta(t - n)
    \label{eq:ap:h_kicked}.
\end{equation}
Its stroboscopic evolution is obtained by integrating the Hamilton equations from just after one kick to just after the next one $t\in[n+\delta, n+1+\delta]$ with $\delta\to 0^{+}$,
which defines the classical map $(q', p') = \M(q, p)$ through
\begin{equation}
    \begin{split}
        q' &= q + g'(p) \\
        p' &= p - f'(q')
    \label{eq:ap:cl_map}.
    \end{split}
\end{equation}
The corresponding quantum map can be obtained by canonical quantization of \eqref{eq:ap:h_kicked} and calculation of the evolution operator for that same stroboscopic evolution (the Floquet operator)
\begin{equation}
    U = e^{-\frac{i}{\hbar}f(\hat{q})} e^{-\frac{i}{\hbar}g(\hat{p})}
    \label{eq:ap:qu_map}.
\end{equation}
Let $z = (q, p)$, given a coherent state $\ket{\alpha}$ with $\alpha = (q + i p)/\sqrt{2\hbar}$
we will calculate the semiclassical expansion of $U^{\dagger}\ket{\alpha}$
in the position representation $\ket{y}$.
By inserting momentum and position identities
$\mathbb{I} =\int d\xi \ketbra{\xi}$, $\mathbb{I} = \int dx \ketbra{x}$
before and after $U^{\dagger}$, respectively, one gets
\begin{equation}
    \mel{y}{U^{\dagger}}{\alpha} =
    \frac{1}{\sqrt{2(\pi\hbar)^{3/2}}}
    \int d\xi e^{\frac{i}{\hbar}(g(\xi) + \xi y)} I_x(\xi)
    \label{eq:ap:semicl_1},
\end{equation}
where
\begin{equation}
    I_x(\xi) = \frac{1}{\sqrt{2\pi\hbar}}
               \int dx\, e^{\frac{i}{\hbar}\phi(x, \xi)} e^{-\frac{1}{2\hbar}(x - q)^2}
    \label{eq:ap:semicl_Ix1},
\end{equation}
with $\phi(x, \xi) = f(x) - \xi x + px$.
The Gaussian factor localizes the integrand around $q$,
which we can make explicit through a change of variables $x = q + \sqrt{\hbar} u$.
Expanding the phase in powers of $\hbar$:
$\phi(x, \xi) = f(q) - \xi q + pq + \sqrt{\hbar}u (f'(q) - \xi + p) + 1/2 f''(q) \hbar u^2 + O(\hbar^{3/2})$,
we can ignore higher order terms since they amount to an additive $O(\sqrt{\hbar})$ correction to $I_x$.
Replacing this expansion back into \eqref{eq:ap:semicl_Ix1} with the change of variables yields
\begin{multline}
    I_x(\xi) = \frac{1}{\sqrt{2\pi}} e^{\frac{i}{\hbar}(f(q) - \xi q + pq)}
    \int du\, e^{\frac{i}{\sqrt{\hbar}}(f'(q) - \xi + p)u} \times\\
    e^{-\frac{1}{2}(1 - if''(q))u^2} + O(\sqrt{\hbar})
    \label{eq:ap:semicl_Ix2}
\end{multline}
The remaining integral is the Fourier transform of a Gaussian evaluated at $(f'(q) - \xi + p)/\sqrt{\hbar}$,
so we finally get
\begin{equation}
    I_x(\xi) = \frac{1}{\sqrt{1 - if''(q)}} e^{\frac{i}{\hbar}(f(q) + pq)}
    e^{-\frac{1}{2\hbar}\frac{(\xi - p - f'(q))^2}{1 - i f''(q)}} + O(\sqrt{\hbar})
    \label{eq:ap:semicl_Ix3}.
\end{equation}
Inserting this back into Eq.~\eqref{eq:ap:semicl_1} yields
\begin{equation}
    \mel{y}{U^{\dagger}}{\alpha} =
    \frac{e^{\frac{i}{\hbar}(f(q) + pq)}}{(\pi\hbar)^{1/4}\sqrt{1 - if''(q)}} I_\xi(y) + O(\sqrt{\hbar})
    \label{eq:ap:semicl_1_2},
\end{equation}
where
\begin{equation}
    I_\xi(y) = \frac{1}{\sqrt{2\pi\hbar}}
            \int d\xi\, e^{\frac{i}{\hbar}(g(\xi) + \xi y - \xi q)}
            e^{-\frac{1}{2\hbar}\frac{(\xi - p - f'(q))^2}{1 - i f''(q)}}
    \label{eq:ap:semicl_Ixi1}.
\end{equation}
Much like before, this integral is localized by the Gaussian (the part with real cuadratic exponent),
and a completely analogous procedure can be utilized to calculate it.
Together with the previous expansion of $f$,
this amounts to considering only up to cuadratic terms in the Hamiltonian \eqref{eq:ap:h_kicked},
or the linearized version of the map \eqref{eq:ap:cl_map}.
The result reads
\begin{equation}
    I_\xi(y) = \frac{\sqrt{1 + i\delta}}{\abs{\sigma}} e^{ \frac{i}{\hbar} \varphi_U(\alpha)}
            e^{-\frac{1}{2\hbar} \frac{1 + i \delta}{\abs{\sigma}^2} (y - q')^2}
            e^{\frac{i}{\hbar}p'y}
         + O(\sqrt{\hbar})
    \label{eq:ap:semicl_Ixi2},
\end{equation}
where we have defined
$\sigma = 1 - g'' f'' + i g''$, $\delta = g'' - f'' + g''f''$
and
$\varphi_U(\alpha) = f + p q + g - p'q$ is a phase dependent on the coherent state and the map,
where $p' \equiv p + f'(q)$, $q' \equiv q - g'(p')$.
In all the previous expressions $f$ and its derivatives are evaluated at $q$,
and $g$ and its derivatives are evaluated at $p'$.
Putting it all together we finally get
\begin{equation}
    \mel{y}{U^{\dagger}}{\alpha} =
    \frac{\sqrt{\sigma}}{\abs{\sigma}}
    \frac{ e^{\frac{i}{\hbar} \varphi_U(\alpha)} }{ (\pi\hbar)^{1/4} }
            e^{-\frac{1}{2\hbar} \frac{1 + i \delta}{\abs{\sigma}^2} (y - q')^2}
            e^{\frac{i}{\hbar}p' y}
         + O(\sqrt{\hbar}).
    \label{eq:ap:semicl_1_3}
\end{equation}
We have thus obtained that a coherent state localized at a point $z=(q, p)$
is evolved in one step of the inversed quantum map \eqref{eq:ap:qu_map}
into a \textit{squeezed} coherent state centred at a point $z'=(q', p')$
given by the inverse of the classical map \eqref{eq:ap:cl_map} $\M^{-1}$ as
\begin{equation}
    \begin{split}
        p' &= p + f'(q) \\
        q' &= q - g'(p')
    \label{eq:ap:cl_inv_map},
    \end{split}
\end{equation}
with a squeezing determined by the second derivatives at that point $f''(q)$, $g''(p')$,
up to an $O(\sqrt{\hbar})$ correction.
From now on we drop the $O$ notation and switch to the notation $\approx$ denoting the lowest-order term in powers of $\hbar$.

All that's left is to compute the projection onto another coherent state
$\mel{\beta}{U^\dagger}{\alpha}$ with $\beta(w) = (x + i\xi)/\sqrt{2\hbar}$
and $w=(x, \xi)$,
which is tedious but straightforward:
insert a position identity $\mathbb{I} = \int dy \ketbra{y}$ after $\bra{\beta}$,
use Eq.~\eqref{eq:ap:semicl_1_3},
expand each quadratic term in $y$ inside the exponentials and complete the squares;
the remaining integral will be the Fourier transform of a Gaussian.
Rearranging things carefully, the result is
\begin{multline}
    \mel{\beta}{U^\dagger}{\alpha} \approx
    \sqrt{\frac{2\sigma}{\abs{\sigma}^2 + 1 + i\delta}}
    e^{\frac{i}{\hbar}(\varphi_U(\alpha) - \vartheta_U(\beta,\alpha))} \times\\
    e^{-\frac{1}{4\hbar}
        (w - z')\mathbb{V}^{-1}(w - z')^t
    }
    \label{eq:ap:beta_U_alpha},
\end{multline}
with the additional phase
\begin{multline*}
\vartheta_U(\beta,\alpha) = \frac{1}{(\abs{\sigma}^2+1)^2 + \delta^2}
\Bigg\{
    \frac{1}{2}\abs{\sigma}^2 \delta
    \Big[  (x-q')^2 - (\xi-p')^2 \Big] \\
    +
    (\xi - p')\Big[\abs{\sigma}^2(\abs{\sigma}^2 + 1)x + (\abs{\sigma}^2+1+\delta^2)q'\Big]
\Bigg\},
\end{multline*}
where the covariance matrix is
\begin{equation}
    \mathbb{V}^{-1} =
    \frac{2}{(\abs{\sigma}^2+1)^2 + \delta^2}
    \begin{pmatrix}
        \abs{\sigma}^2 + 1 + \delta^2 & - \abs{\sigma}^2 \delta \\
        - \abs{\sigma}^2 \delta & \abs{\sigma}^2(\abs{\sigma}^2 + 1)
    \end{pmatrix}
    \label{eq:ap:cov_matrix},
\end{equation}
which is a proper non-singular and positive-definite matrix for all values of $f'', g''$
with determinant $\det(\mathbb{V}^{-1}) = 4\abs{\sigma}^2/[(\abs{\sigma}^2+1)^2 + \delta^2]$.
Finally, we obtain
\begin{equation}
    \frac{1}{2\pi\hbar}\abs{\mel{\beta(w)}{U^\dagger}{\alpha}}^2
    \approx
    \frac{1}{2\pi\hbar\sqrt{\det(\mathbb{V})}}
    e^{-\frac{1}{2\hbar} (w - z')\mathbb{V}^{-1}(w - z')^t}
    \label{eq:ap:abs_beta_U_alpha_2},
\end{equation}
a 2D Gaussian distribution.
By diagonalizing $\mathbb{V} = U^{-1} \mathbb{D} U$ (where $U$ is an orthogonal matrix)
and changing variables $\zeta = (w - z')U$ (note that $d\zeta = dw$ since $\det(U) = 1$)
this can be rewritten as two separate, independent 1D Gaussian distributions,
each with zero mean and variance directly proportional to $\hbar$:
$$
    \frac{1}{2\pi\hbar}\abs{\mel{\beta(\zeta U^{-1} + z')}{U^\dagger}{\alpha}}^2
    \approx
    \mathcal{N}(u|0, \hbar\lambda_1) \mathcal{N}(v|0, \hbar\lambda_2),
$$
where $\lambda_1$ and $\lambda_2$ are $\mathbb{V}$'s eigenvalues
and $\zeta = (u, v)$.
The $0$-variance limit of a Gaussian distribution is a Dirac delta,
so taking the $\hbar\to 0$ limit yields
$$
\frac{1}{2\pi\hbar}\abs{\mel{\beta(\zeta U^{-1} + z')}{U^\dagger}{\alpha}}^2
\approx
\delta(u)\delta(v) = \delta(\zeta),
$$
and changing back to the original variables we finally arrive at
\begin{equation}
    \frac{1}{2\pi\hbar}\abs{\mel{\beta(w)}{U^\dagger}{\alpha(z)}}^2
    \approx
    \delta(w - \M^{-1}(z))
    \label{eq:ap:abs_beta_U_alpha_delta},
\end{equation}
after using that $\delta(zU) = \delta(z)/\det(U) = \delta(z)$ for any $z$,
and recalling that $z' = \M^{-1}(z)$ [Eq.~\eqref{eq:ap:cl_inv_map}].
Replacing Eq.~\eqref{eq:ap:abs_beta_U_alpha_delta} into Eq.~\eqref{eq:ap:husimi_Urho}:
$$
    H_{\U \rho}(z) \approx P_{\rho}(\M^{-1}(z)).
$$
Finally, Eq.~\eqref{eq:ap:hus_P} allows us to write
\begin{equation}
    P_{\U \rho}(z) \approx P_\rho(\M^{-1}(z)) \equiv \PF P_\rho(z)
    \label{eq:ap:P_semicl_final},
\end{equation}
where we have used the definition of the Perron-Frobenius operator [Eq.~\eqref{eq:perron_frobenius}].

This calculation was carried out for a single application of the map,
and we don't prove the more general statement
$$
    P_{\U^t \rho}(z)
    \approx \PF^t P_\rho(z).
$$
At a fixed value of $\hbar$, this will hold for as long as there is a semiclassical correspondence,
which is the Ehrenfest time $t \leq \tau_E$.
This timescale is controlled by the stability of the classical trajectory
such that $\tau_E \sim 1/\hbar^{1/2}$ for a stable trajectory
and $\tau_E \sim \abs{\log(\hbar)}/\lambda$ for an unstable one~\cite{Shepelyansky:2020, PhysRevLett.89.040403},
where $\lambda$ is the Lyapunov exponent.

As a final note,
we haven't considered maps defined on the torus for this calculation,
but since for small enough values of $\hbar$
(large enough values of $N$, the dimension of the Hilbert space for the quantized torus)
the discrepancies become vanishingly small, the same result holds.


\section{Lack of correspondence with alternative methods}
\label{sec:ap:lack_correspondence}
In the main text we see quantum-to-classical correspondence of the Arnoldi procedure,
and thus of the Krylov states and complexity,
for the von Neumann evolution of a density matrix.
This is, its quantum Krylov space converges to the purely classical Krylov space for the distribution that defines its Glauber-Sudarshan P-representation \eqref{eq:p_representation}.
One alternative approach may be to apply the Arnoldi iteration to the evolution of a
pure density matrix $\ketbra*{\alpha(x)}$ or even a ket $\ket*{\alpha(x)}$.
In both cases, the corresponding Husimi distribution is a Gaussian
centred at the point $x=(q,p)$ with variance $\sigma^2 = \hbar$.
Thus in the classical limit $\hbar\to 0$ one may expect
the result to match that of the classical evolution of the same Gaussian distribution for sufficiently small timescales~\cite{PhysRevA.71.010101, PhysRevE.79.025203, PhysRevA.65.042113}.
In this appendix we point out that this is not a viable alternative in the current context.

\begin{figure}[ht]
    \vspace{0.1cm} 
    \centering
    \includegraphics[width=\linewidth]{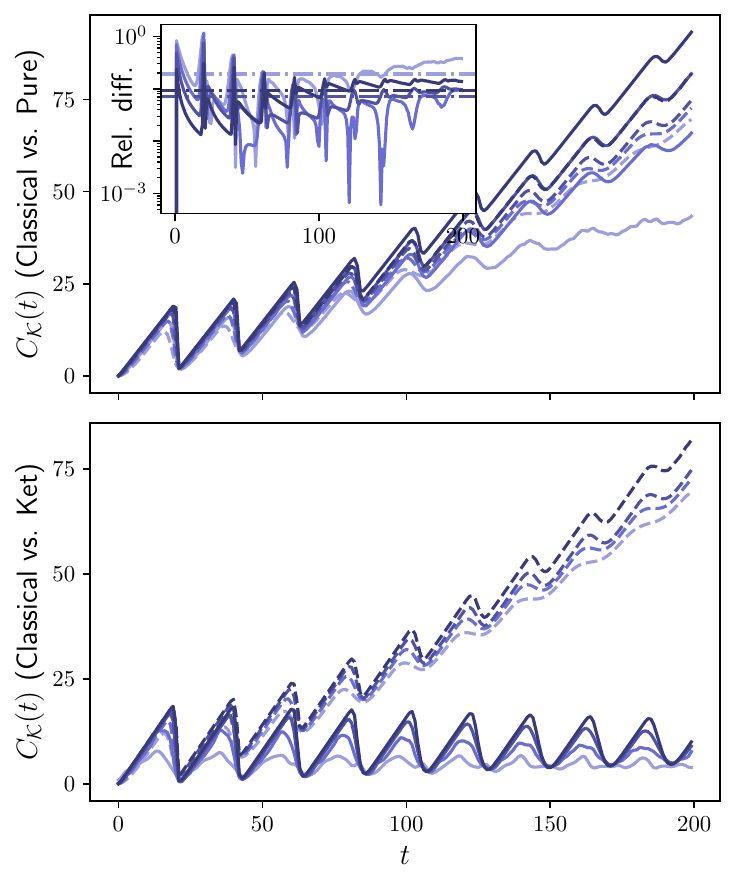}
    \caption{Comparison of the Krylov complexity for the quantum evolution of
        a pure coherent-state density matrix (top, solid lines),
        a coherent state (bottom, solid lines)
        and the classical (dashed lines) evolution of a Gaussian distribution with variance
        $\sigma^2 = \hbar = 1/2\pi N$ for the Harper map, with $N\in \{2^6, 2^7, 2^8, 2^9\}$ (light to dark).
        The inset shows the relative difference (to the classical) between the classical and quantum complexities (solid lines) and their average values (dash-dotted lines).}
    \label{fig:ap:Harper_ket_pure_complexity}
\end{figure}
\begin{figure}[ht]
    \centering
    \includegraphics[width=\linewidth]{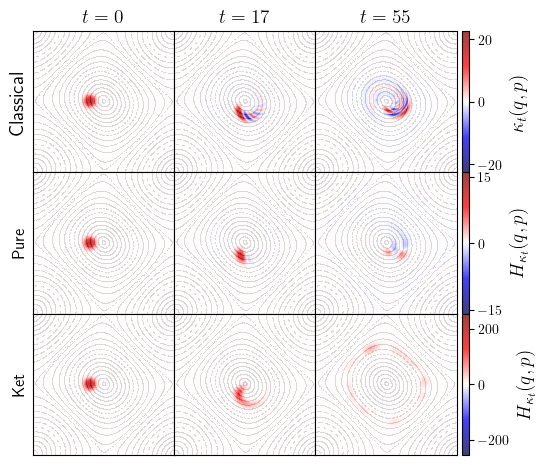}
    \caption{Comparison of the Krylov states generated at various times $t$
        for the quantum evolution of a pure coherent density matrix,
        a coherent state and the classical evolution of a Gaussian distribution with variance
        $\sigma^2 = \hbar = 1/2\pi N$ for the Harper map, with $N=2^8$.
        The quantum Krylov states are represented via the Husimi distribution [Eq.~\eqref{eq:husimi}].
    }
    \label{fig:ap:Harper_ket_pure_states}
\end{figure}

In Fig.~\ref{fig:ap:Harper_ket_pure_complexity}
we show comparisons between the Krylov complexity obtained for the evolution of
an initially pure coherent-state density matrix, a coherent state, and a classical Gaussian distribution with variance $\sigma^2 = \hbar$, as discussed just above, for the Harper map.
Figure \ref{fig:ap:Harper_ket_pure_states} shows some of the corresponding Krylov states.
Aside from the variance, all other parameters have been kept the same as in the main text.
The Krylov complexity for the coherent state is completely different to the classical
(note that the Krylov complexity of states is also referred to as the spread complexity),
and while that of the density matrix does fare better,
there is no clear correspondence.
In fact, the average relative difference between the quantum and classical \textit{increases}
with a decrease in $\hbar$ when going from $N = 2^8$ to $N= 2^9$ [Fig.~\ref{fig:ap:Harper_ket_pure_complexity}].

Compared to those of the classical evolution,
the states generated by the density matrix are only quantitatively different,
while those of the state are also \textit{qualitatively} different, having a distinct structure altogether
[Fig.~\ref{fig:ap:Harper_ket_pure_states}].
One may expect the latter two to show some structural differences
since one is an operator Krylov space constructed from the propagator $\U$ with inner product~\eqref{eq:product_q},
and the other a state space from the usual unitary evolution operator $U$ with the usual ket inner product.
The space of density matrices is not closed upon the operations carried out in the Arnoldi iteration,
and as a result the initially pure state generates Krylov states that are not pure.
On the other hand, the space of kets \textit{is} closed, so pure states remain pure,
and the Husimi distributions of the resulting Krylov states are by necessity positive.
These two observations encapsule why the Krylov state of the pure density matrix is not equivalent to that of the ket.

Note that, while the profiles in phase space of the initial Krylov states are identical in the three cases, the scales are completely different (see the colorbars in Fig~\ref{fig:ap:Harper_ket_pure_states}).
This is set by the values of the packets at their center, which in these examples are
$1/\sqrt{\pi\hbar}$, $1/\sqrt{2\pi\hbar}$ and $1/2\pi\hbar$ for the distributions corresponding to the classical, the pure state and the ket, respectively.
This could in principle be ``corrected" by introducing an $\hbar$-dependent factor in the definition of the inner product, which does not alter the functional form of the Krylov states.

The aforementioned differences in the complexities can be interpreted from the distinct features of the generated Krylov states.
The states generated by the pure density matrix have suppressed tails
(compared to the classical)
meaning most of the weight in \eqref{eq:kn_rhot} is set on the coefficient accompanying $\rho_n$,
which in turn implies a large overlap $\sbraket{\kr_n}{\rho_t} \equiv \bnt{n}{t}$ for $n \approx t$,
and as a result the Krylov complexity~\eqref{eq:k_complexity} is consistently higher for the pure density matrix
as seen in Fig.~\ref{fig:ap:Harper_ket_pure_complexity}.
One striking feature of the Krylov states generated by the coherent state $\ket*{\alpha(x)}$
is that beyond the first period $T\sim 25$
they are localized \textit{outside} the torus
[See Fig.~\ref{fig:ap:Harper_ket_pure_states}].
Because after the period the newly generated states lie beyond the torus,
the evolution has little overlap with them $\bnt{n}{t} \sim 0$ for $n \gtrsim T$
and the wavefunction is confined to the interval $n\in[0, T]$.
The oscillations in the Krylov complexity decay, and it settles into a value $\Ck \sim T/2$
as the wavefunction becomes uniform in that interval.


\bibliography{references}

\begin{thebibliography}{53}
\bibitem [{asy()}]{asymptoticNotation}%
  \BibitemOpen
  \href@noop {} {}\bibinfo {note} {We denote by $X \approx Y$ that $Y$ is the
  lowest-order term in the asymptotical expansion of $X$ in terms of
  $\hbar$.}\BibitemShut {Stop}%
\end{thebibliography}
\end{document}